# Fundamental Concepts of Cyber Resilience: Introduction and Overview

Igor Linkov, U.S. Army Engineer Research and Development Center

Alexander Kott, U.S. Army Research Laboratory

## Motivation: Why Cyber Resilience?

Society is increasingly reliant upon complex and interconnected cyber systems to conduct daily life activities. From personal finance to managing defense capabilities to controlling a vast web of aircraft traffic, digitized information systems and software packages have become integrated at virtually all levels of individual and collective activity. While such integration has been met with immense increases in efficiency of service delivery, it has also been subject to a diverse body of threats from nefarious hackers, groups, and even state government bodies. Such cyber threats have shifted over time to affect various cyber functionalities, such as with Direct Denial of Service (DDoS), data theft, changes to data code, infection via computer virus, and many others.

Attack targets have become equally diverse, ranging from individuals to international companies and national government agencies. At the individual level, thousands of personal data records including credit card information and government identification is stolen on a daily basis – disrupting the lives of many persons and generating billions of dollars in fraud or other losses. At the corporate level, hacking attempts targeted at the Sony Corporation, Equifax, and other similarly sized organizations demonstrate the potential for hackers to gain entry to sensitive information stored in company databases, and potentially impact the security of millions of users. Lastly, state-based cyber threats arise from individual hackers and other large states alike, such as with daily intrusion attempts that occur within the Department of Defense. While many cyber threats are thwarted, many are able to exact lasting and widespread damage in terms of security, financial losses, social disorder, and other concerns. In warfare, cyber threats may soon become one of the main factors that decide whether a war is won or lost (Kott et al 2015).

Whereas traditional risk assessment comprises a calculation of product of threats, vulnerabilities, and consequences for hazards and their subsequent exposures, risk assessment becomes limited in the cybersecurity field as approaches are needed to address threats and vulnerabilities that become integrated within a wide variety of interdependent computing systems and accompanying architecture (DiMase et al. 2015; Ganin et al. 2017). For highly complex and interconnected systems, it becomes prohibitively difficult to conduct a risk assessment that adequately accounts for the potential cascading effects that could occur through an outage or loss spilling over into other systems. Given the rapid evolution of threats to cyber systems, new management approaches are needed that address risk across all interdependent domains (i.e., physical, information, cognitive, and social) of cyber systems (Linkov et al. 2013a, b).

Further, the unpredictability, extreme uncertainty, and rapid evolution of potential cyber threats leaves risk assessment efforts all the more unable to adequately address cybersecurity concerns for critical infrastructural systems. For this reason, the traditional approach of hardening of cyber systems against identified threats has proven to be impossible. The only true defense that cybersecurity professionals could take to harden systems from the multitude of potential cyber threats would include the disallowance of cyber systems from accessing the internet. Therefore, in the same way that biological systems develop immunity as a way to respond to infections and other attacks, so too must cyber systems adapt to ever-changing threats that continue to attack vital system functions, and to bounce back from the effects of the attacks (Linkov et al. 2014).

For these reasons, cyber resilience refers to the ability of the system to prepare, absorb, recover and adapt to adverse effects, especially those associated with cyber-attacks. (We will discuss the exact definitions later.) Here, depending of the context, we use the term cyber resilience to refer mainly to the resilience property of a system or network; sometimes we also use the term as referring to the features or components of the system that enable cyber resilience.

## Resilience and Systems

Cyber resilience should be considered in the context of complex systems that comprise not only physical and information but also cognitive and social domains (Smith, 2005). Cyber Resilience ensures that system recovery occurs by considering interconnected hardware, software and sensing components of cyber infrastructure (Fig 1). It is thus constitute a bridge between sustaining operations of the system while ensuring mission execution.

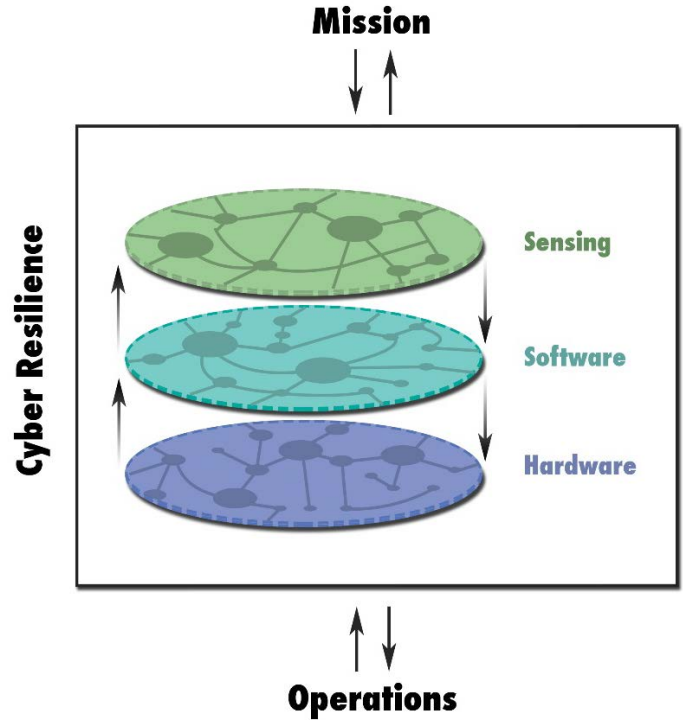

Figure 1. The cyber resilience domains comprise sensing, hardware, and software components which collectively contribute to sustaining system operations. .

.

Resilience has roots in many disciplines and integrates ecological, social, psychological, organizational, and engineering perspectives and definitions. Resilience engineering, for example, has been defined as "the ability of systems to anticipate and adapt to the potential for surprise and failure," and has been associated with a shift in safety paradigm acknowledging that system coping is important when prevention is impossible (Hollnagel, Woods, & Leveson, 2006). Ecological resilience, on the other hand, refers to the ability of the system to absorb and withstand shocks, with an emphasis on persistence (Holling, 1996). Resilience is used as a metaphor to describe how systems react to stressors, and to bridge the gap in understandings between fields, resilience needs to be discussed less abstractly, separating the metaphor from the science. Across the many diverse lines of inquiry, there are weak linkages between concepts and methods for resilience.  Useful ideas and results accumulate and partially overlap but it is often difficult to find the common areas.  In addition, the different technical languages hamper communication of ideas about resilience across the different contributing disciplines and application problems.

Despite multi-disciplinary nature of resilience and multiple definition, there are common themes and resilience features across these multiple disciplines (Connolly et al (2017). Resilience defined by the National Academies of Science (NAS) as "the ability to prepare and plan for, absorb, recover from, and more successfully adapt to adverse events" is emerging as one of the most widely used by various organizations and governance agencies (Larkin et al., 2015).  The common resilience features include critical functions (services), thresholds, cross-scale (both space and time) interactions, and memory and adaptive management. The concept of *critical functionality* is important to understanding and planning for resilience to some shock or disturbance. *Thresholds* play a role in whether a system is able to absorb a shock, and whether recovery time or alternative stable states are most salient. *Recovery time* is essential in assessing system resilience after a disturbance where a threshold is not exceeded. Finally, the concepts of *memory* describes the degree of self-organization in the system, and *adaptive management* provides an approach to managing and learning about a system's resilience opportunities and limits, in a safe-to-fail manner. Connelly et al., 2017 related these features to the National Academy of Sciences definition of resilience (Table 1), including the temporal phases of the NAS definition to emphasizing the importance of time in all conceptualizations of resilience.

**Table 1.** Resilience features common to socio-ecology, psychology, organizations, and engineering and infrastructure, which are related to the temporal phases from the National Academy of Sciences definition of resilience (from Connelly et al., 2017, with permission).

| NAS phase of resilience | Resilience Feature | Description by Application Domain | | | |
|---|---|---|---|---|---|
| | | Socio-Ecological | Psychological | Organizational | Engineering & Infrastructure |
| *Plan* | **Critical Functions (Services)** | A system function identified by stakeholders as an important dimension by which to assess system performance | | | |
| | | Ecosystem services provided to society | Human psychological well-being | Goods and services provided to society | Services provided by physical and technical engineered systems |
| *Absorb* | **Thresholds** | Intrinsic tolerance to stress or changes in conditions where exceeding a threshold perpetuates a regime shift | | | |
| | | Used to identify natural breaks in scale | Based on sense of community and personal attributes | Linked to organizational adaptive capacity and to brittleness when close to threshold | Based on sensitivity of system functioning to changes in input variables |
| *Recover* | **Time (and Scale)** | Duration of degraded system performance | | | |
| | | Emphasis on dynamics over time | Emphasis on time of disruption (i.e., developmental stage: childhood vs adulthood) | Emphasis on time until recovery | Emphasis on time until recovery |
| *Adapt* | **Memory/Adaptive Management** | Change in management approach or other responses in anticipation of or enabled by learning from previous disruptions, events, or experiences | | | |

| | | Ecological memory guides how ecosystem reorganizes after a disruption, which is maintained if the system has high modularity | Human and social memory, can enhance (through learning) or diminish (e.g., post-traumatic stress) psychological resilience | Corporate memory of challenges posed to the organization and management that enable modification and building of responsiveness to events | Re-designing of engineering systems designs based on past and potential future stressors |
|---|---|---|---|---|---|

## Resilience and Related Properties of Systems

Similarly to other fields, cyber resiliency refers to the system's ability to recover or regenerate its performance after a cyber attacks produces a degradation of its performance. For now, until we delve further into metrics of cyber resilience, we can say the following: assuming two equally performing systems  A and B subjected to an impact (resulting from a cyber-attack) that left both systems with an equal performance degradation, the resiliency of system A is greater if after a given period T it recovers to a higher level of performance than that of system B.

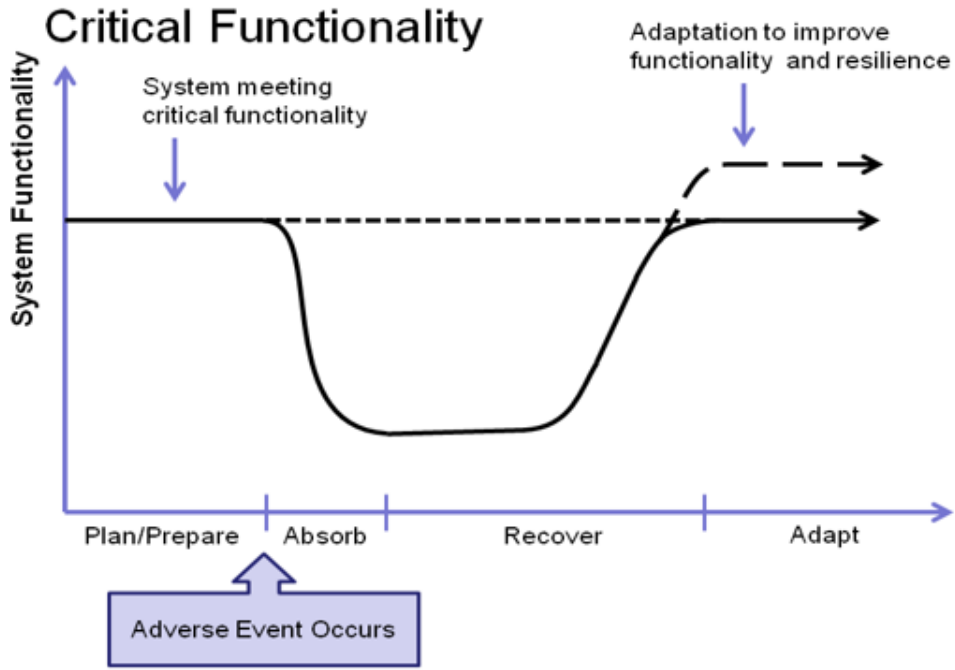

Resilience is often confused or conflated with several other related but different concepts. These include risk, robustness, and security. Oxford dictionary clearly defined these concepts. Risk is "a situation involving exposure to danger [threat]." If risk is managed appropriately, system reaches state of security (i.e. "the state of being free from danger or threat") or robustness ( i.e "the ability to withstand or overcome adverse conditions or rigorous testing). Security, robustness and risks are connected, they are focused on preventing system from degrading and keeping functionality within acceptable level before and after the adverse event. Resilience is a very different concept. Oxford defines it as "the capacity to recover quickly from difficulties." Resilience assessment thus starts with the assumption that system is affected, functionality is impaired and is focused on evaluation of recovery speed.

The literature on cyber risk (including here what some call "IT risk") most commonly defines cyber risk in terms that combine likelihood of an undesirable event, and a measure of the impact of the event. Although several approaches to risk assessment exist, the methods adopted by US regulatory agencies are largely based around the risk = threat $x$ vulnerability $x$ consequence model. For example, NIST's description from NIST Publication SP 800-30 (NIST) states: "Risk is a function of the likelihood of a given threat-source's exercising a particular potential vulnerability, and the resulting impact of that adverse event on the organization. To determine the likelihood of a future adverse event, threats to an IT system must be analyzed in conjunction with the potential vulnerabilities and the controls in place for the IT system." ISO's definition of IT risk is similar: the potential that a given threat will exploit vulnerabilities of an asset or group of assets and thereby cause harm to the organization. It is measured in terms of a combination of the probability of occurrence of an event and its consequence (ISO/IEC 2008).

The key components of cyber risk are relatively well understood. The likelihood of a successful cyber attack can be empirically measured and estimated apriori with a degree of accuracy from known characteristics of a system or network (Leslie et al 2017), (Gil et al 2014). The cyber impact on a system is a topic in which assessment methods are being developed (Kott et al 2017). Since cyber threats are difficult to quantify, current efforts shift from quantifying risk in specific units (like probability of failure) towards risk-based decision making using multi-criteria decision analysis (Ganin et al., 2017). Unlike the concept of resilience, the concept of risk does not answer the questions of how well the system is able to absorb a cyber attack, or how quickly and how completely the system is able to recover from a cyber attack. Whenever risks are identified and actions taken to reduce risk, there still remains residual risk. As such, resilience assessment and management is, in part, an effort to address that remaining known, but unmitigated, risk as well as enhance the overall ability of the system to respond to unknown or emerging threats.

Robustness is another concept often confused with resilience. Robustness is closely related to risk. Robustness denotes the degree to which a system is able to withstand an unexpected internal or external threats or change without degradation in system's performance. To put it differently, assuming two systems – A and B—of equal performance, the robustness of system A is greater than that of system B if the same unexpected impact on both systems leaves system A with greater performance than system B. We stress the word *unexpected* because the concept of robustness focuses specifically on performance not only under ordinary, anticipated conditions (which a well-designed system should be prepared to withstand) but also under unusual conditions that stress its designers' assumptions. For example, in IEEE Standard 610.12.1990, "Robustness is defined as the degree to which a system operates correctly in the presence of exceptional inputs or stressful environmental conditions." Similarly, "robust control refers to the control of unknown plants with unknown dynamics subject to unknown disturbances" (Chandrasekharan 1996). Note that the length of recovery typically depends on the extent of damage. There may also be a point beyond which recovery is impossible. Hence, there is a relation between robustness (which determines how much damage is incurred in response to an unexpected disturbance) and resiliency (which determines how quickly the system can recover from such damage). In particular, a system that lacks robustness will often fail beyond recovery, hence offering little resiliency. Both robustness and resiliency, therefore, must be understood together.

## Making Business Case for Cyber Resilience

Traditional risk assessment is appealing for cyber risk governance due to the quantitative nature and the single risk value that is output. These characteristics make risk thresholds easy to formalize in policy documents and to regulate in a consistent manner at the federal level. However, quantitative risk assessment typically involves quantification of the likelihood of an event's occurrence and of the vulnerability to the event. Emerging cyber realities and technologies are presenting new threats with uncertain intensity and frequency and the vulnerabilities and consequences in terms of the extent of casualties, economic losses, time delays, or other damages are not yet fully understood or modeled. As a result, risk calculations become more uncertain and generate costly solutions since multiple, often hypothetical, threat scenarios could point to many vulnerabilities and catastrophic system failures that are unaffordable to mitigate, absorb, or recover. Furthermore, users and other stakeholders may have preferences for accepting some loss in performance of one part of the system over any degradation in another part (Bostick, Holzer, and Sarkani 2017). One outcome can be significant funding spent in ways that do not align with stakeholder values, resulting in dissatisfaction with performance, despite the expense, or even litigation that interferes with the risk reduction efforts.

A key risk management strategy is to identify critical components of a system that are vulnerable to failure and subsequently, to harden them. This approach can be appropriate for many isolated cyber systems, but when the nature of the threat is unknown, as discussed above, it is difficult to identify all of the critical components and it becomes increasing expensive to act conservatively to harden or protect all parts of the system against all types of threats (Figure 1). The result has been stagnation in investment. As risk mitigation plans become more expensive and are delayed while funding is sought, infrastructure and societal systems are left largely unprepared for emerging and uncertain threats (Meyer 2011). Furthermore, there are fewer and fewer isolated systems in our world and the degree of interdependency and interconnectedness can be difficult to characterize and quantify.

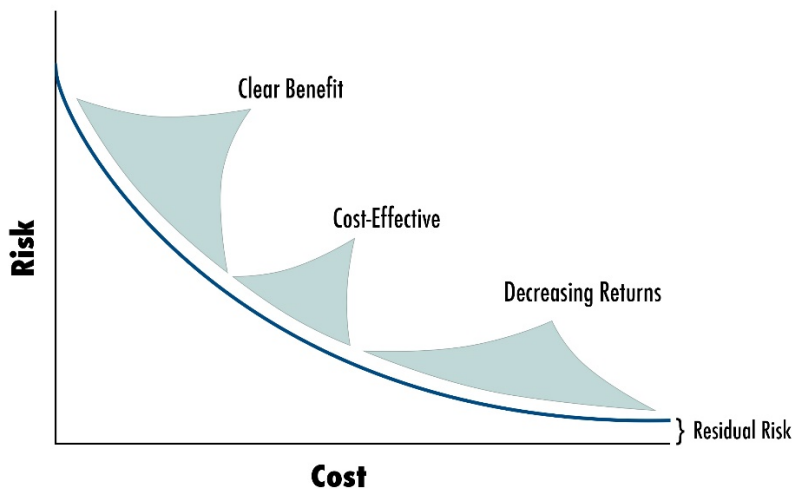

**Figure 1. Conceptual diagram of the cost of buying-down risk in cyber systems.**

In exchange for accepting the current levels of risk rather than demanding greater preventative and protective measures, funding can be re-allocated towards resilience enhancement efforts. For systems that have already completed cost-effective risk reduction measures, Figure 3 shows the funding that can be re-allocated toward resilience by accepting risk level (a) over risk level (b). In parallel to these public changes, the academic community should be called on to develop decision models that identify the optimal investment in risk reduction versus resilience and recovery improvement.

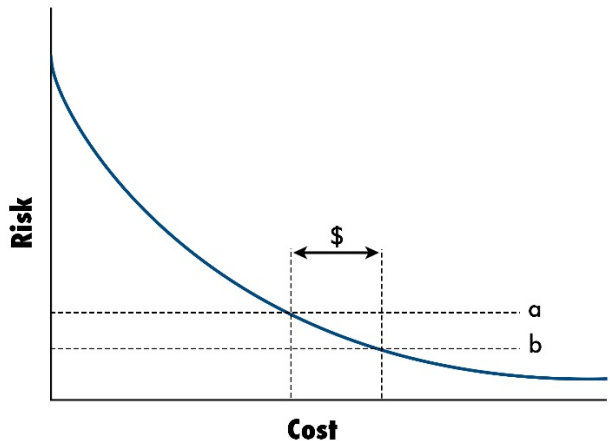

**Figure 3. Existential threat and mega-cost avoided when resources are re-allocated to resilience—e.g., accepting risk level (a) instead of risk level (b).**

## Assessment of Cyber Resilience

Myriad tools and methods marketed as resilience assessments now exist, but take very different formats (Linkov and Florin, 2016; Nordgren, Stults and Meerow, 2016; Arnott, Moser, and Goodrich, 2016). Some are as simple as a checklist, others are geo-spatial visualizations of quantifiable metrics, while still others are network modeling methods but with no generalized form, they are custom built for each application. The outputs of these tools are similarly varied, including maps, scores, and process time graphs. Developers of the tools span a wide range of entities including academic, private (e.g., consulting), program sponsors (e.g., foundations and agencies), boundary organizations that bridge across research, policy, and practitioner realms, and potentials users themselves. Potential users include state and city managers, industry process administrators, and utility operators, many lacking the expertise to choose among the rapidly accumulating products in this emerging field.

Figure 2 shows two primary approaches currently described in the literature to address resilience, including metric-based and model-based approaches. Metric-based approaches use measures of individual properties of system components or functions to assess overall system performance while  model-based approaches use system configuration modeling and scenario analysis to predict system evolution. In general, metrics are defined as measurable properties of the system that quantify the degree to which the objectives of the system are achieved. Metrics provide vital information pertaining to a given system, and are generally acquired by way of analyzing relevant attributes of that system. Some researchers and practitioners make a distinction between a measure and a metric, whereas others may refer to metrics as performance measures (Collier et al 2016). Resilience metrics are discussed in detail in (Linkov et al 2013b), and two chapters of this book are dedicated to exploring alternative approaches to defining cyber resilience metrics.

Despite differing levels of mathematical complexity, the scientific challenges are in addressing system-level processes and tailoring methodologies to specific needs. A number of efforts have been focused on developing metrics that are applicable to a variety of systems, including social, ecological and technical (Eisenberg et al. 2014). The current lack of universally applicable resilience metrics as well as inability to formalize value systems relevant to the problems at hand have been barriers to wide implementation of metric-based methodologies. Advances in decision analysis and social and economic valuation of benefits offer ways to address these challenges, with methods to assess the impact of trading off resilience attributes (e.g., flexibility, redundancy) with values currently considered in the decision-making process (e.g., cost, environmental impact, risk reduction) for diverse investment alternatives. Further research on this topic can greatly benefit both management and investment decisions for system resilience.

Model-based approaches focus on a representation of the real world and a definition of resilience using mathematical or physical concepts. Modeling requires knowledge of the critical functions of a system, mission, the temporal patterns of a systems, thresholds, and system memory and adaptation. Process models require a detailed understanding of the physical approaches within a system to simulate event impacts and system recovery and are difficult to construct and are information consuming. Statistical approaches alternatively require a lot of data on system performance. Bayesian models combine features of process and statistical models. Network models require a presentation of the system as interconnected networks whose structure is dependent on the function of the system. Alternatively, the game theoretical/agent-based approach focuses on the model performance of the system based on a limited set of rules defined by modelers. Using these approaches, resilience can be defined but the utility of many advanced models is limited because of the data intensive requirements. Network science is emerging as an important tool to allow quantitative framing for the future of resilience as a scientific discipline. In network science, the system is represented as an interconnected network of links and nodes that exhibit behavior in space and time. These methods have been demonstrated; though only for limited case studies where network recovery was explicitly modeled (Ganin et al. 2016; Gao, Barzel, and Barábasi 2016; Cohen et al. 2000). The challenge is to frame resilience as characteristic of several major network properties that would provide a universal foundation to the field with cross-domain applications, similar to the threat-vulnerability-consequence framework used in the field of risk analysis. The four parameters of resilience (critical function, thresholds, time, and memory) will be the basis of identifying and describing the relevant network properties. This shift in thinking and assessment tools is needed to encourage adaptability and flexibility in addition to adequately assessing the tradeoffs between redundancy and efficiency that characterize a useful resilience assessment.

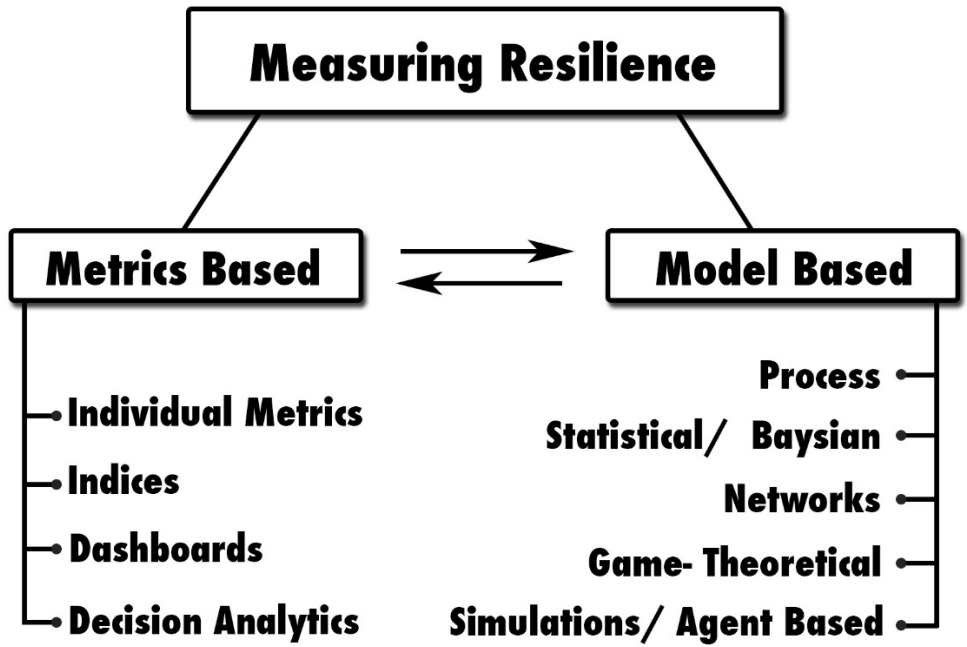

Figure 2.Metric-based and model-based approaches for Resilience Assessment. Multiple tools have been developed to address resilience in systems in both methodological groups.

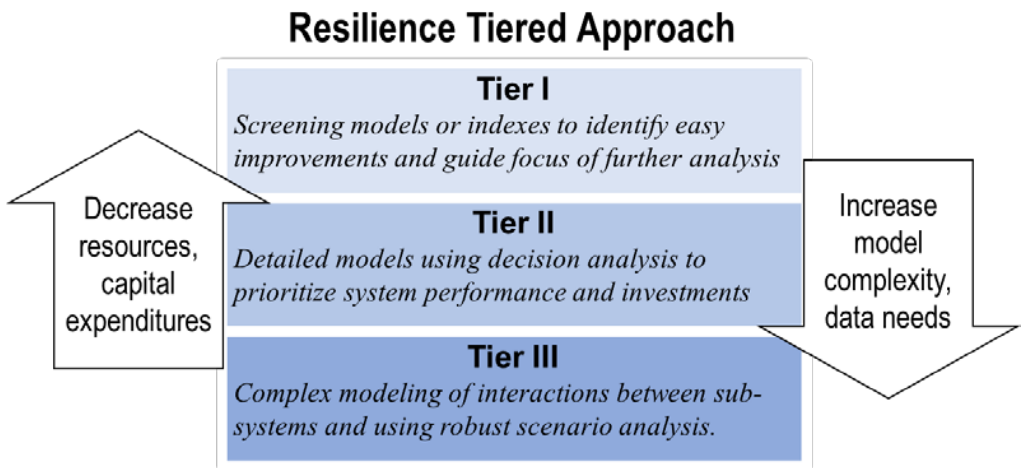

Figure 3. Overview of tiered approach to resilience assessment

Linkov et al (2018) proposed combining all multiple tools for resilience assessment in a tired framework (Figure 1). The goal of each tier is to describe the performance and relationship of critical systems in order to identify management options that enhance performance in parallel with activities that reduce risk. The methods of Tier I quickly and inexpensively identify the broad functions that a system provides to human society or the environment and prioritize the performance of the critical functions both during and in the time following a disruptive event. Analytically, this framing and characterization analysis makes use of existing data, expert judgment, and conceptual models. The methods of Tier II describe the general organization and relationships of the system in a simple form such as a process model or critical path model. Identifying sequential and parallel events in a disturbance can reveal feedback processes and dependencies that are the root of cascading system failures. The methods of Tier III build a detailed model of important functions and related sub-systems where each process and each component of the system is parameterized. The process can be halted at any tier when enough information has been synthesized such that actionable system investments or projects to improve system resilience, given available resources, have been identified by the decision makers.

## Approaches to Improving Cyber Resilience

Resilience of a system, a network or an organization is influenced by a number of factors, in a complex and often contradictory manner. In this section, we consider some of these factors, and how they can be managed or exploited in order to enhance the resilience. In addition to this section, further in this book, the chapter by Keys et al discusses general practices, and the chapter by Bodeau and Graubart describes a set of frameworks, analytic methods, and technologies that can be used to improve system and mission cyber resilience.

**Manage Complexity**: Resilience of a system or network depends greatly on complexity of links within the system (Kott and Abdelzaher 2014).  In his pioneering work, Perrow (1984) explains that catastrophic failures of systems emerge from high complexity of links which lead to interactions that the system's designer cannot anticipate and guard against. System's resiliency precautions can be defeated by hidden paths, incomprehensible to the designer because the links are so numerous, heterogeneous, and often implicit.

This issue is particularly important in multi-genre networks, which are networks that combine several distinct genres – networks of physical resources, communication networks, information networks, and social and cognitive networks. When we consider an entire multi-genre network—and not merely one of the heterogeneous, single-genre sub-networks that comprise the whole – we see far more complexity of the paths connecting the network's elements.

Of particular importance are those paths within the system that are not recognized or comprehended by the designer. Indeed, the designer can usually devise a mechanism to prevent a propagation of failure through the links that are obvious. Many, however are not obvious, either because there are simply too many paths to consider – and the numbers rapidly increase once we realize that the paths between elements of a communication system, for example, may also pass through a social or an information network—or because the links are implicit and subtle. Subtle feedback links may lead to a failure in organizational decision making (Kott 2006).

To enhance resilience, in some cases, the designer of the system can use greater complexity of connections between two elements of the systems by increasing redundancy of its functions. Also, as the number and heterogeneity of links grow, they offer the agents (or other active mechanisms) within the network more opportunities to regenerate the network's performance. These agent may be able to use additional links to more elements in order to reconnect the network, to find replacement resources, and ultimately to restore its functions.

On the other hand, greater complexity of the network may also reduce the resiliency of the network. For example, active agents may be more likely to be confused by the complexity of the network, or to be defeated in their restoration work by un-anticipated side effects induced by hidden paths within the network. The increase in complexity may also lead to lower resilience by increasing—and hiding from the designer—the number of ways in which one failed component may cause the failure of another. Therefore, in most cases greater complexity should be avoided when possible, unless it directly supports resilience functions. In this book, the chapter by Evans and Horsthemke explains the role of dependencies in complex systems and how to analyze and characterize the impact of dependencies; in another chapter they provide an example of analysis of a large-scale, highly complex web of systems called regional critical infrastructure. The chapter by Bodeau and Graubart describe the techniques of segmentation and isolation that can be used to manage the complexity in order to enhance resilience.

**Chose Topology**: Quite apart from complexity, the choice of appropriate topology of the system or network can improve resilience. Much prior research addressed the fundamental vulnerabilities of different networks as a function of their topological properties. Of particular interest has been the classification of properties of networks according to their node degree distribution. While some networks (such as wireless and mesh networks) are fairly homogeneous and follow an exponential node degree distribution, others, called scale-free networks (such as the Web or the power-grid), offer significant skew in node degrees, described by a power-law. It is well known that scale-free graphs are much more robust to random node failures (errors) than graphs with an exponential degree distribution, but that these scale-free graphs are increasingly more vulnerable to targeted attacks (namely, removal of high-degree nodes). In this book, the chapter by Moore and Cho explores the role of topology and methods to analyze the influence of topology on resilience; and the chapter by Kotenko et al presents topology-based methods in analysis of cyber attack propagation and the impact on resilience.

**Add Resources**: additional resources in a network can help improve resiliency. For example, adding capacity to nodes in a power generation and distribution network may reduce likelihood of cascading failures and speed up the service restoration. Adding local storage and influencing the distribution of nodes of different functions in a network also leads to improved resilience at the expense of additional resources. Resiliency may be improved by adding multiple functional capacities to each node (usually implying the need for additional resources), by processing more input sources (requiring more resources for acquisition of inputs and for processing), or by combination of multiple parallel processing mechanisms. Yet the same measures tend to increase complexity and might cause greater difficulties in restoring the network's capability if degraded by an unexpected -- and probably harder to understand--failure.

Providing redundant resources can help both to absorb and to restore the system. Redundancy, however, should be used with caution. If the designer adds identical redundant software or hardware, the same malware would be able to compromise multiple redundant resources. If diversity is introduced, and the redundant resources are significantly different, the complexity grows with its potential negative impact on resilience. In this book, the chapter by Musman offers an example of estimating mission resilience in comparing two options of adding resources: (a) adding a replicate server or (2) adding a fast recovery resource. The chapter by Bodeau and Graubart discusses how technologies and processes for contingency planning and COOP, including Diversity and Redundancy, support resilience. The chapter by Curado et al describes the Fog Services concept in which functions are widely distributed over large number of resources.

**Design for Reversibility**: Components of the system should be designed in a manner that allows them to revert to a safe mode when failed or compromised. This means several things: (1) the component in the failed mode should not cause any further harm to itself or other components of system or environment; (2) it should be possible to reverse the state of the component in the process of recovering the system. This is because some failures, such as physical breakage and human losses are often irreversible or expensive to remedy (e.g., once there is a reactor meltdown, it is hard to "roll back"). This characteristic is unlike purely logical systems (e.g., databases), where roll-back from failure is more feasible and cheaper.

Note however, that conventional fail-safe design practice could be incompatible with need for the system to absorb the failure, and therefore may reduce resilience. For example, the operator of a system notices that a computer is compromised by a malware. A reasonable fail-safe action might be to disconnect the computer. However, this might be detrimental to the overall resilience of the system if the computer is needed to support other components that execute damage absorbing actions.

**Control Propagation**: To enhance the system's ability to absorb the impact of a cyber compromise, the designer should guard against cascaded failures. Such failures are non-independent in that one triggers another. A network that is prone to large "domino effects" will likely sustain severe damage in response to even modest disturbances, which significantly limits the scope within which efficient absorption and recovery (and hence resilient operation) remains possible. Therefore, the dependencies or links between nodes should designed, whenever possible, in a way that minimizes the likelihood that a failure propagates from easily from one node to another. Ideally, links should both passively and actively filter out the propagation of failure. One possible form of such filtering is buffering, discussed next in this section. Further in this book, the chapter by Moore and Cho investigates propagation of failures and cascading failures. The chapter by Giacomello and Pescaroli discusses cascading failures and the role of human factors in propagation of failure.

**Provide Buffering:** In data and commodity flow networks, the function of the network is to offer its clients access to a set of delivered items. In such networks, buffers (e.g., caching, local storage) constitute a resiliency mechanism that obviates the need for continued access to the original source. Should the original source become unreachable, one can switch to a local supplier. Hence, local access can be ensured despite interruption of the global supply network as long as access to a local cache (buffer) is available. Local access is an especially valuable solution in the case of a data flow network, where the commodity (namely, the data content) is not consumed by user access, in the sense that a local distributer can continue to serve a content item to new users irrespective of its use by others. Much work on network buffering has been done to increase the resiliency of data access to fluctuations in resource availability. For example, buffering (or caching) has been used to restore connectivity and performance upon topology changes in ad hoc networks, as well as to reduce access latency in disruption-tolerant network.

**Prepare Active Agents:** Active agents – human or artificial -- should be available to take active measures in order to absorb, recover and adapt. In order to be effective in doing so, the agents must have plans, processes and preparation. Where appropriate and necessary, human operators or users of the system should play the role of active resilience agents. Wherever possible, however, the designer of the system should consider introducing into the system a set of artificial partly-autonomous intelligent agents that are able to conduct the absorption and recovery actions, in an autonomous manner (Kott et al 2018).

In order to perform the absorb and recover actions in presence of the adversary malware deployed on the friendly system, the artificial agent often has to take destructive actions, such as deleting or quarantining certain software. Such destructive actions are carefully controlled by the appropriate rules of engagement. Developers of the agent should attempt to design its actions and planning capability to minimize the associated risk. This risk has to be balanced against the destruction caused by the adversary if the agent's action is not taken. Because the adversary and its malware know that the agent exists and is likely to be present on the system, the malware seeks to find and destroy the agent. Therefore, the agent possesses techniques and mechanisms for maintaining a degree of stealth, camouflage and concealment. More generally, the agent takes measures that reduce the probability that the malware will detect the agent. The agent is mindful of the need to exercise self-preservation and self-defense.

When humans are the active resilience agents, in order to be effective these human agents must be appropriately trained, prepared, motivated. They should have skills, resources and processes available to them, to perform the functions of absorb, recover and adapt. The human organization should be properly structured, roles and responsibilities clearly defined, collective skills developed, and team training and rehearsals conducted.

In this book, a number of chapter focus on various related aspects. The chapter by Key et al discusses organizational measures and human resources practices. The chapter by Bodeau and Graubart mention non-persistence, realignment, and adaptive responses. The chapter by Evans talks about active measures including MTD and Cyber Deception. The chapter by Giacomello and Pescaroli explores human factors and organizational culture. And the chapter by Tandiya et al presents and compares AI techniques that might be considerate for implementing response strategies.

**Build Agents Capabilities**: Ideally, agents should be able to perform one of multiple functions depending on context, and the same function could be performed by one of several agents. For example, storage agents (buffers, caches) in a network can use their space to store any of a set of possible items. Also, the same item can be stored by any of multiple agents. For example, individuals in an organization can allocate their time to any of a set of possible projects. Similarly, the same project can be performed by any of multiple individuals. The combination of versatility and redundancy of agents significantly improves resiliency of network functions by facilitating reconfiguration to adapt to perturbations. Intuitively, the higher the versatility of the individual agents and the higher the degree to which they are interchangeable, the more resilient is the system to perturbation because it can reallocate functions to agents more flexibly to restore its performance upon resource loss. Another useful dimension is agent's capacity; that is, the number of different functions that an agent can simultaneously perform. Capacity quantifies, for example, the number of items a storage node can simultaneously hold, or the number of projects a given individual can simultaneously work on. On a related note, in this book, Curado et al describe a distributed decision mechanism supported by multiple SDN controllers intended to enhance recovery mechanisms.

**Consider Adversary**: If the adversary specifically tailors his techniques and procedures – and possesses the necessary capabilities – in order to defeat specifically the absorption and recovery efforts, the systems resilience will suffer accordingly. The designer of the system should consider the likely adversary's capabilities, intent, tactics, techniques and procedures, and design the mechanisms and processes of absorption and recovery in a manner that are more likely to withstand the adversary actions. Game-theoretic analysis and wargaming – manual or computerized – should be conducted in order to optimize the proposed measures (Colbert et al 2017). In this book, the chapter by Kotenko et al considers explicit modeling of adversaries attack methods. The chapter by Bodeau and Graubart discusses how resilience-enhancing measures may need to be adapted in case of an adversary that constitutes an Advanced Persistent Threats.

**Conduct Analysis**: As noted in several places in this section, all resilience-enhancing measures and actin can also cause un-anticipated effects leading to overall reduction in resilience. Therefore, rigorous, high-fidelity analysis is a must. A resilience-enhancing measure should not be designed or introduced without an appropriate analysis that is capable of revealing potential negative impacts and systemic effects (Kott et al 2017). Comparative analytical studies should be conducted with and without the proposed measure. High-fidelity, simulation-based analysis is particularly appropriate. The fidelity of the simulation should be sufficient to replicate multiple modes of propagation, modes of interactions, feedback channels, and second and third-order effects. Because adversary actions and counteractions play a great role in cyber resilience, the analysis must include the adversary as well.  In this book, the chapter by Musman offers recommendations for conducting analysis that leads to estimating a mission resilience metrics.  The chapter by Ormrod and Turnbull reviews topics related to simulation of complex systems for analysis of resilience. The chapter by Rose et al addresses economic aspects in comparative analysis of different resilience-enhancing techniques. The chapter by Karsai et al describes an example of a toolkit for simulation-based resilience analysis.

## Preview of the Book

Reflecting the key themes we have covered in the introduction, the first three parts of the book cover the topics of quantification, assessment and analysis, and enhancement of cyber resilience. The fourth, final, part is dedicated to cases studies of selected classes of systems and networks.

The first part presents two alternative – but not incompatible – views on how to quantify cyber resilience via suitable metrics. It opens with a chapter by Cybenko that takes the perspective in which system performance is central to the metrics. As discussed in the Introduction chapter of this book, cyber resiliency has become an increasingly important, relevant and timely research and operational concept in cyber security. Although multiple metrics have been proposed for quantifying cyber resiliency, a connection remains to be made between those metrics and operationally measurable and meaningful resilience concepts that can be empirically determined in an objective manner. This chapter describes a concrete quantitative notion of cyber resiliency that can be tailored to meet specific needs of organizations seeking to introduce resiliency into their assessment of their cyber security posture.

If the previous chapter showed how to quantify cyber resilience from the perspective of system performance, the chapter by Musman et al takes an alternative view – the perspective of mission risk. The chapter describes the features that any definition of resilience should consider to enable measurable assessment and comparison. It proposes a definition of resilience that supports those considerations. Then, it reviews and discusses in detail the terminology and definitions that have been proposed in the context of these considerations. Ultimately, the chapter chooses a definition of resilience that relates to "mission risk." When based on risk, the authors of this chapter argue, a resilience definition is clearly defined, measurable, and has a sound theoretical grounding. Since risk relies on both the likelihood of events occurring as well as changes in value (i.e., damage) when these events occur, we are provided with a computable metric that enables assessment and comparison. This allows us to tailor metrics to specific systems.

The second part of the book focuses on approaches to assessment and analysis of cyber resilience. Having discussed, in the previous two chapters, perspectives on quantifying cyber resilience, we now present several chapters that assemble qualitative and quantitative inputs for a broad range of metrics that might apply to cyber resilience. Some of these approaches (e.g., most of this chapter and the next one) are largely qualitative and based on human review and judgement of pertinent aspects of systems, organization and processes. Other are based on quantitative and often theoretically rigorous modeling and simulation of systems, networks and processes.

The purpose of the chapter by Keys and Shapiro is to outline best practices in an array of areas related to cyber resilience. While by no means offering an exhaustive list of best practices, the chapter provides an organization with means to "see what works" at other organizations. It offers these best practices within existing frameworks related to dimensions of cyber resilience. The chapter begins with a discussion of several existing frameworks and guidelines that can be utilized to think about cyber resilience. Then, the chapter describes a set of "best practices" based on a selection of metrics from these frameworks. These best practices can help an organization as a guide to implementing specific policies that would improve their cyber resilience.

The general overview of frameworks and best practices of cyber resilience assessments provided in the previous chapter is followed by the chapter by Evans and Horsthemke that focuses more specifically on methodologies that use the concept of cyber dependencies. A cyber dependency is a connection between two components, such that these components' functions depend on one another and loss of any one of them degrades the performance of the system as a whole. Such dependencies must be identified and understood as part of a cyber resilience assessment. This chapter describes two related methodologies that help identify and quantify the impact of the loss of cyber dependencies. One relies on a facilitated survey and dependency curve analysis, and helps an organization understand its resiliency to the loss of a dependency. That methodology incorporates the ability of an organization to withstand a loss through backup (recovery) methods and assess its resiliency over time. Another methodology helps an organization consider the indirect dependencies that can cause cascading failures if not sufficiently addressed through protective measures. However, that methodology does not incorporate protective measures such as redundancy or consider the possibility that the loss of a dependency might not have an immediate impact.

Unlike the previous chapter where propagation of failures along the dependency links was studied in a qualitative, human-judgement fashion, the chapter by Moore and Cho offers an approach to analyzing resilience to failure propagation via a rigorous use of percolation theory. In percolation theory, the basic idea is that a node failure or an edge failure (reverse) percolates throughout a network and, accordingly, the failure affects the connectivity among nodes. The degree of network resilience can be measured by the size of a largest component (or cluster) after a fraction of nodes or edges are removed in the network. In many cybersecurity applications, the underlying ideas of percolation theory have not been much explored. In this chapter, it is explained how percolation theory can be used to measure network resilience in the process of dealing with different types of network failures. It introduces the measurement of adaptability and recoverability in addition to that of fault-tolerance as new contributions to measuring network resilience by applying percolation theory.

The chapter by Kotenko and co-workers continues exploring how resilient is a network to a failure propagating through it; however, now we also include an explicit treatment of specific causes of failure – malicious activities of the cyber attacker. This chapter considers cyber-attacks and the ability to counteract their implementation as the key factors determining the resilience of computer networks and systems. Indeed, cyber-attacks are the most important among destabilizing forces impacting a network. Moreover, the term cyber resilience can be interpreted as the stability of computer networks or systems operating under impact of cyber-attacks. The approach in this chapter involves the construction of analytical models to implement the most well-known types of attacks. The result of the modeling is the distribution function of time and average time of implementation of cyber-attacks. These estimates are then used to find the indicators of cyber resilience. To construct analytical models of cyber-attacks, this chapter introduces an approach based on the stochastic networks conversion, which works well for modeling multi-stage stochastic processes of different nature.

So far, the discussion was limited to relatively narrow abstractions of systems and networks. Such narrow abstractions allow effective assessment and analysis methodologies, but do not cover the richness and diversity of realistic organizations, systems and processes. Therefore, the chapter by Ormrod and Turnbull explains how to build a multi-dimensional simulation model an organization's business processes. This multi-dimensional view incorporates physical objects, human factors, and time and cyberspace aspects. Not all systems, the components within a system, or the connections and interfaces between systems and domains are equally resilient to attack. It is important to test complex systems under load in a variety of circumstances to both understand the risks inherent in the systems, but also to test the effectiveness of redundant and degenerate systems. There is a growing need to test and compare the limitations and consequences of potential mitigation strategies before implementation. Simulation is a valuable tool because it can explore and demonstrate relationships between environmental variables in a controlled and repeatable manner. This chapter introduces the Integrated Cyber-Physical Effects (ICPE) model as a means of describing the synergistic results obtained through the simultaneous, parallel or sequential prosecution of attacking and defensive measures in both the physical and cyber domains.

Suppose you assessed or analyzed the resilience of a system using approaches described in Part 2 of this book, or similar approaches. Chances are, you determined that the resilience of the system is inadequate, at least in part. Would should you do to improve it? This is the theme of Part 3 of this book: methods, techniques and approaches to enhancing cyber resilience of a system, either via an appropriate initial design, or by adding mitigation measures, or by defensive actions during a cyber-attack.

The chapter by Bodeau and Graubart opens the theme with a broad overview of approaches to enhancing systems resilience in the spirit of Systems Engineering. It starts by providing background on the state of the practice for cyber resilience. Next, the chapter describes how a growing set of frameworks, analytic methods, and technologies can be used to improve system and mission cyber resilience. For example, technologies and processes created for contingency planning and COOP can be adapted to address advanced cyber threats. These include Diversity and Redundancy. Cybersecurity technologies and best practices can be extended to consider advanced cyber threats. These include Analytic Monitoring, Coordinated Defense, Privilege Restriction, Segmentation, and Substantiated Integrity.

In the previous chapter, we were introduced to active defense among numerous other approaches. Now, in the chapter by Evans, we explore active defense techniques in detail. These are automated and human-directed activities that attempt to thwart cyber-attacks by increasing the diversity, complexity, or variability of the systems and networks. These limit the attacker's ability to gather intelligence, or reduces the usable lifespan of the intelligence. Other approaches focus on gathering intelligence on the attackers, either by attracting attackers to instrumented honeypots or by patrolling the systems and networks to hunt for attackers. The intelligence gathering approaches rely upon cyber security personnel using semi-automated techniques to respond and repel attackers. Widely available commercial solutions for active defense so far are lacking. Although general purpose products may emerge, meanwhile organizations need to tailor their applications for available solutions, or develop their own customized active defense. A successfully architected system or application should include passive defenses, which add protection without requiring human interaction, as well as active defenses.

Technology solutions have been our focus so far. Now, the chapter by Giacomello and Pescaroli notes the possibility that the human component of critical infrastructures, instead of the mere technological one, could be the primary vector of events constituting less than resilient behavior of a system. This chapter introduces a systemic approach that contextualizes cascading dynamics in the vulnerability of their technological as well as human assets. It is followed by a wider focus on the evolution of critical infrastructure and management, envisioned as root causes of cascades, introducing the role of the human factor in that process. The chapter highlight why any investment in technological resilience of cyber assets cannot be missing the integration of its human component, defining practical suggestions for the field. In fact, there is a growing consensus among security experts is that the weakest link in the security chain is the human being, whether as users, customers, administrators or managers. The technological progress needs to be followed step by step by improvement in operator's skills and routines, adjusting their improvisational behaviors and resilience.

The next chapter, by Linkov and co-workers, continues the topic we started to discuss in the previous chapter – the human factors. However, it focuses on a specific method of enhancing cyber resilience via establishing appropriate rules for employees of an organization under consideration. Such rules aim at reducing threats from, for example, current or former employees, contractors, and business partners who intentionally use their authorized access to an organization to harm the organization. System users can also unintentionally contribute to cyber-attacks, or themselves become a passive target of a cyber-attack. The implementation of work-related rules is intended to decrease such risks. However, rules implementation can also increase the risks that arise from employee disregard for rules. This can occur when the rules become too restrictive, and employees become more likely to disregard the rules. Furthermore, the more often employees disregard the rules both intentionally and unintentionally, the more likely insider threats are able to observe and mimic employee behavior. This chapter shows how to find an intermediate, optimal collection of rules between the two extremes of 'too many rules' and 'not enough rules'.

Recent years have seen continuous, rapid growth in popularity and capabilities of Artificial Intelligence, and broader speaking, of other computational techniques inspired by biological analogies. It is most appropriate, therefore, for this book to explore how such techniques might contribute to enhancing cyber resilience.  The chapter by Tandiya and co-workers argues that the fast-paced development of new cyber-related technologies complicates the classical approach of designing problem-specific algorithms for cyber resilience. Instead, "general-purpose" algorithms – such as biologically-inspired artificial Intelligence (BIAI) -- are more suited for such problems. BIAI techniques allow learning, adaptability, and robustness, which are compatible with cyber resilience scenarios like self-organization, dynamic operation conditions, and performance in adversarial environment. The chapter introduces the readers to BIAI techniques and describes various BIAI techniques and their taxonomy. It also proposes metrics which can be used to compare the techniques in terms of their performance, implementation ease, and requirements. Finally, the chapter illustrates the potential of such techniques via several case studies -- applications pertaining to wireless communication systems.

Implementation of means for enhancing cyber resilience – any of the means we have discussed in the preceding chapters -- costs money. Is this a worthwhile investment? The chapter by Rose and co-workers provides economic perspective on how to choose the most economically-appropriate approaches to improving cyber resilience. These considerations are rather complex. For example, property damage, except for destruction of data, has thus far been a relatively minor cost of cyber threats, in contrast to instances of significant loss of functionality of a cyber system itself or the system it helps operate.  The latter translates into loss of output (sales revenue and profits) and loss of employment, and is often referred to as business interruption (BI).  Thus, post-disaster strategies that enable a system to rebound more efficiently and quickly hold the prospects of greatly reducing BI. Moreover, there are numerous resilience tactics on both the cyber service provider side and customer side, many of which are relatively inexpensive. The latter include back-up data storage and equipment, substitutes for standard cyber components, conserving on cyber needs, and recapturing lost production once the cyber capability is restored.  The chapter describes the analysis based on basic principles of economics and is couched in a benefit-cost analysis (BCA) framework as an aid to decision-making.  The chapter goes beyond the conceptual level and offers estimates of the costs and effectiveness of various mitigation and resilience tactics.

The chapter by Evans and Horsthemke opens the last, fourth, part of our book. In this part we explore several cases where cyber resilience was addressed – from different perspectives – in application to complex systems or networks. We collected these cases to answer the question of a practically-minded reader, "How do I approach assessing of enhancing resilience of a particular system I am interested in?" While these few cases cannot cover all possible classes of systems or networks, they serve as useful illustrative examples and could inform approaches to resilience in other classes of systems.

The opening chapter of this part discusses large-scale, highly complex web of systems called regional critical infrastructure. These are responsible for providing entire large regions (the size of states or countries) with water, electricity, natural gas, communications, transportation, healthcare, police, fire protection, and ambulances. Often, these are further complicated by multiple, not always fully cooperative owners of these systems; and by the diversity of threats that may attacks such systems, ranging from natural disasters to state-sponsored cyber attackers. The chapter focuses mainly on particular cyber resiliency assessment (CRA) methodology. The foundation of the methodology are collective assessments by human stakeholders and experts, seeking areas of concerns and developing options for resilience enhancements. CRA involves analytical and modeling techniques for cyber assessments. The chapter illustrates the application of CRA to a regional critical infrastructure with a realistic case study.

Then Curado and co-workers address cyber-physical systems resilience with a focus on Internet of Things (IoT) as a particularly prominent example of large-scale cyber-physical systems. The emphasis is on current and future network architectures and systems, highlighting main research issues and technological trends. The chapter opens by discussing and contrasting resilience of organization and resilience of communications and computing technologies. It then proceeds to explore issues of resilience in two use cases. One case deals with smart cities (arguably a form of IoT) and another with large scale networks. The chapter points out that the Internet of Things is evolving towards an Internet of Everything, where everybody and everything are connected to provide multiple services within various contexts such as Smart Home, Wearables, Smart City, Smart Grid, Industrial Internet, Connected Car, Connected Health, Smart Retail, Smart Supply Chain and Smart Farming. In the context of this evolution, a number of challenges must be addressed, most of which touch on issues of resilience, among others.

Pacheco and co-workers continue the discussion of cyber-physical systems, including Internet of Things, with a special focus on resilient services for Smart Cities. The topic of Smart Cities has been already introduced in the previous chapter, and now, this chapter presents a detailed approach to design and development of resilient services for Smart Cities based on Moving Target Defense (MTD) and autonomic computing paradigms. Moving Target Defense is often seen as a game changing approach to building self-defending systems. In an earlier chapter "Active Defense Techniques" MTD has been introduced in a broader context. In the specific instantiation of MTD in this chapter, it dynamically randomizes the resources used, and the execution environment to run CPS applications, so that the attackers (outsiders or insiders) cannot determine the resources used to run the provided services and consequently are unable to evade attacks. The chapter discusses both the detailed methodology for applying MTD to enhance resilience of Smart, and the experimental results obtained with implemented prototypes.

A transportation network is a critical component of a Smart City (considered in the preceding chapter), and therefore it is fitting that a distinguishing element of the next chapter, by Karsai and co-workers, is the resilience analysis of transportation networks. The chapter highlights the importance of humans in most cyber-physical systems and uses the term Human Cyber Physical System (H-CPS). It further argues that H-CPS design processes should use five fundamentally different abstraction layers: the Physical Layer, the three "cyber" layers: Network, Service Platform and Application Layers, and the Human Layer. It them describes describe the Cyber-Physical Systems Wind Tunnel (CPSWT), a simulation integration architecture and tool kit, and proceeds to illustrate a simulation-based resilience analysis using a transportation network example.

Supply chains are among the most exposed and vulnerable component of any system. The last chapter of the book, by Collier and co-workers, explores the resilience perspective of supply chain and begins by identifying a set of factors that enable resilience. It also explains the nature of actors within the supply chain and discusses possible metrics for characterizing cyber resilience of supply chains, as well as of broader systems in which a supply chain is a component. To a large extent, this chapter – and in fact the entire book -- is a review of proposed research agenda on many topics that are yet to be addressed in our quest to understand the ways of quantifying, analyzing and enhancing the cyber resilience.

## References

Bostick, T. P., Holzer, T. H., & Sarkani, S. (2017). Enabling stakeholder involvement in coastal disaster resilience planning. *Risk Analysis, 37*(6), 1181–1200.

Bostick, T. P., Connelly, E. B., Lambert, J. H., & Linkov, I. (2018). Resilience Science, Policy and Investment for Civil Infrastructure. Reliability Engineering & System Safety *175*:19–23.

Chandrasekharan, P. C. (1996). *Robust control of linear dynamical systems*. London: Academic Press.

Cohen, R., Erez, K., Ben-Avraham, D., & Havlin, S. (2000). Resilience of the internet to random breakdowns. *Physical Review Letters, 85*(21), 4626.

This is a preprint version of the chapter that appears in the book "Cyber Resilience of Systems and Networks," Springer 2018

Colbert, E. J., Kott, A., Knachel III, L., & Sullivan, D. T. (2017). *Modeling Cyber Physical War Gaming* (Technical Report No. ARL-TR-8079). US Army Research Laboratory, Aberdeen Proving Ground, United States.

Collier, Z. A., Linkov, I., DiMase, D., Walters, S., Tehranipoor, M., & Lambert, J. (2014a). Risk-Based Cybersecurity Standards: Policy Challenges and Opportunities. Computer 47:70–76.

Collier, Z. A., Walters, S., DiMase, D., Keisler, J. M., & Linkov, I. (2014b). A semi-quantitative risk assessment standard for counterfeit electronics detection. *SAE International Journal of Aerospace, 7*(1), 171–181.

Collier, Z. A., Panwar, M., Ganin, A. A., Kott, A., & Linkov, I. (2016). Security metrics in industrial control systems. In *Cyber-security of SCADA and other industrial control systems* (pp. 167–185). Cham: Springer International Publishing.

Connelly, E. B., Allen, C. R., Hatfield, K., Palma-Oliveira, J. M., Woods, D. D., & Linkov, I. (2017). Features of resilience. *Environment Systems and Decisions, 37*(1), 46–50.

DiMase, D., Collier, Z. A., Heffner, K., & Linkov, I. (2015). Systems engineering framework for cyber physical security and resilience. *Environment Systems and Decisions, 35*(2), 291–300.

Eisenberg, D. A., Linkov, I., Park, J., Bates, M., Fox-Lent, C., & Seager, T. (2014). Resilience metrics: Lessons from military doctrines. *Solutions, 5*(5), 76–87.

Florin, M. V., & Linkov, I. (Eds.), (2016) *IRGC Resource Guide on Resilience*. Lausanne: EPFL International Risk Governance Council (IRGC).

Ganin, A. A., Massaro, E., Gutfraind, A., Steen, N., Keisler, J. M., Kott, A., Mangoubi, R., & Linkov, I. (2016). Operational resilience: Concepts, design and analysis. *Scientific Reports, 6*, 19540.

Ganin, A. A., Quach, P., Panwar, M., Collier, Z. A., Keisler, J. M., Marchese, D., & Linkov, I. (2017a). Multicriteria decision framework for cybersecurity risk assessment and management. *Risk Analysis*.

Ganin, A., Kitsak, M., Marchese, D., Keisler, J., Seager, T., & Linkov, I. (2017b). Resilience and efficiency in transportation networks. Science Advances 3 (12): e1701079.

Gao, J., Barzel, B., & Barabási, A. L. (2016). Universal resilience patterns in complex networks. *Nature, 530*(7590), 307–312.

Gil, S., Kott, A., & Barabási, A. L. (2014). A genetic epidemiology approach to cyber-security. *Scientific Reports, 4*, 5659.

Holling, C. S. (1996). Engineering resilience versus ecological resilience. In P. C. Schulze (Ed.), *Engineering within ecological constraints*. Washington, D.C.: National Academy Press.

Hollnagel, E., Woods, D. D., & Leveson, N. C. (2006). *Resilience engineering: Concepts and precepts*. Aldershot: Ashgate.

ISO/IEC. (2008). Information technology – Security techniques-Information security risk management ISO/IEC FIDIS 27005, ISO/IEC.

Kaplan, S., & Garrick, B. J. (1981). On the quantitative definition of risk. *Risk Analysis, 1*(1), 11–27.

Kelic, A., Collier, Z. A., Brown, C., Beyeler, W. E., Outkin, A. V., Vargas, V. N., Ehlen, M. A., Judson, C., Zaidi, A., Leung, B., & Linkov, I. (2013). Decision framework for evaluating the macroeconomic risks and policy impacts of cyber attacks. *Environment Systems & Decisions, 33*(4), 544–560.

Kott, A. (2006). *Information warfare and organizational decision-making*. Artech House, Boston, USA.

Kott, A., & Abdelzaher, T. (2014). Resiliency and robustness of complex systems and networks. *Adaptive Dynamic and Resilient Systems, 67*, 67–86.

Kott, A., Alberts, D. S., & Wang, C. (2015). Will cybersecurity dictate the outcome of future wars? *Computer, 48*(12), 98–101.

Kott, A., Ludwig, J., & Lange, M. (2017). Assessing mission impact of cyberattacks: Toward a model-driven paradigm. *IEEE Security and Privacy, 15*(5), 65–74.

Kott., et al. (2018). *A Reference Architecture of an Autonomous Intelligent Agent for Cyber Defense* (Technical Report). US Army Research Laboratory, Aberdeen Proving Ground, United States.

Larkin, S., Fox-Lent, C., Eisenberg, D. A., Trump, B. D., Wallace, S., Chadderton, C., & Linkov, I. (2015). Benchmarking agency and organizational practices in resilience decision making. *Environment Systems and Decisions, 35*(2), 185–195.

Leslie, N. O., Harang, R. E., Knachel, L. P., & Kott, A. (2017). Statistical models for the number of successful cyber intrusions. *The Journal of Defense Modeling and Simulation, 15*(1), 49–63.

Linkov, I., Eisenberg, D. A., Bates, M. E., Chang, D., Convertino, M., Allen, J. H., Flynn, S. E., & Seager, T. P. (2013a). Measurable resilience for actionable policy. *Environmental Science and Technology, 47*(18), 10108–10110.

Linkov, I., Eisenberg, D. A., Plourde, K., Seager, T. P., Allen, J., & Kott, A. (2013b). Resilience metrics for cyber systems. *Environment Systems and Decisions, 33*(4), 471–476.

Linkov, I., Bridges, T., Creutzig, F., Decker, J., Fox-Lent, C., Kröger, W., Lambert, J. H., Levermann, A., Montreuil, B., Nathwani, J., Renn, O., Scharte, B., Scheffler, A., Schreurs, M., Thiel-Clemen, T., & Nyer, R. (2014). Changing the resilience paradigm. *Nature Climate Change, 4*(6), 407–409.

Linkov, I., Fox-Lent, C., Allen, C. R., Arnott, J. C., Bellini, E., Coaffee, J., Florin, M. -V., Hatfield, K., Hyde, I., Hynes, W., Jovanovic, A., Kasperson, R., Katzenberger, J., Keys, P. W., Lambert, J.H., Moss, R., Murdoch, P. S., Palma-Oliveira, J., Pulwarty, R. S., Read, L., Sands, D., Thomas, E. A., Tye, M. R., & Woods, D. (In press). Tiered Approach to Resilience Assessment. Risk Analysis, DOI: 10.1111/risa.12991.

Marchese, D., Reynolds, E., Bates, M. E., Morgan, H., Clark, S. S., & Linkov, I. (2018). Resilience and sustainability: Similarities and differences in environmental management applications. *Science of the Total Environment, 613*, 1275–1283.

Meyer, T. (2011). Global public goods, governance risk, and international energy. *Duke Journal of Comparative & International Law, 22*, 319–347.

NIST. (2012). SP 800–30 Risk Management Guide for Information Technology Systems.

Nordgren, J., Stults, M., & Meerow, S. (2016). Supporting local climate change adaptation: Where we are and where we need to go. *Environmental Science & Policy, 66*, 344–352.

Perrow, C. (1984). *Normal accidents: Living with high risk technologies*. Princeton University Press, Princeton, New Jersey.

Roege, P. E., Collier, Z. A., Mancillas, J., McDonagh, J. A., & Linkov, I. (2014). Metrics for energy resilience. *Energy Policy, 72*(1), 249–256.

Roege, P. E., Collier, Z. A., Chevardin, V., Chouinard, P., Florin, M. V., Lambert, J. H., Nielsen, K., Nogal, M., & Todorovic, B. (2017). Bridging the gap from cyber security to resilience. In I. Linkov & J. M. Palma-Oliveira (Eds.), *Resilience and risk: Methods and application in environment, cyber, and social domains* (pp. 383–414). Dordrecht: Springer.

Smith, E. A. (2005). Effects based operations. Applying network centric warfare in peace, crisis, and war. Command and Control Research Program (CCRP), Office of the Assistant Secretary of Defense, Washington DC.